\documentstyle[prl,aps,multicol,epsf,epsfig]{revtex}
 \newcommand \be {\begin{equation}}
\newcommand \ee {\end{equation}}
 \newcommand \ba {\begin{eqnarray}}
\newcommand \ea {\end{eqnarray}}
\newcommand {\bea} {\begin{eqnarray} \nonumber }
\newcommand {\eea} {\end{eqnarray}}
 \newcommand {\eps} {\epsilon}

\def\(({\left(}
\def\)){\right)}
\def\[[{\left[}
\def\]]{\right]}

\def \intq{\int {d^3q \over (2 \pi)^3}}

\def\bi{\bibitem}

\newcommand {\vu}{{\vec u}}

\def \vr{\vec{r}}
\def \vr1{{\vec r_1}}
\def \vr2{{\vec r_2}}
\def \vu1{{\vec u_1}}
\def \vu2{{\vec u_2}}
\def \vu3{{\vec u_3}}
\begin{document}
\draft
\tightenlines

\title{\bf Pulling Hairpinned Polymers }
\author{ Andrea Montanari$^*$ and Marc M\'ezard $^{**}$}

\address{$^{*}$ Scuola Normale Superiore and INFN -- Sez. di Pisa}
\address{I-56100 Pisa, Italy}

\address{$^{**}$ Laboratoire de Physique Th\'eorique et
Mod\`eles Statistiques}
\address{Universit\'e Paris Sud, Bat. 100, 91 405
Orsay {\sc cedex}, France}

\date{\today}
\maketitle

\begin{abstract}
We introduce a model of thermalized conformations in space of
 RNA -or single stranded DNA- molecules,  which includes the possibility
of hairpin formation. This model  contains the usual
secondary structure information, but extends it to the study
of one element of the ternary structure, namely the end-to-end distance. 
 The computed
force-elongation characteristics is in good agreemement with some recent
measurements on single stranded DNA molecules.
\end{abstract}

Recent progress in the manipulation of single biomolecules is making
gradually accessible a wealth of interesting physical information. 
One of the basic investigations concerns the 
force-elongation characteristics: its measurement in double
stranded DNA (dsDNA) molecules has provided very interesting 
results in the last few years,
going from a detailed characterization of the elastic properties of the molecules
to the existence of new phases of dsDNA in various regimes of tension and
overcoiling
\cite{smith,perk,strick,marsig,sDNA,smith2,strick2,strick3,pDNA,heslot,gaub}.

While the force-elongation characteristics of dsDNA is rather 
well understood, the corresponding knowledge on single stranded DNA (ssDNA)
is poorer:
although in some ionic conditions it may be characterized by a simple
freely jointed chain (FJC) with elastic bonds \cite{smith2}, this description is 
not valid when one changes the ionic concentrations
\cite{maier}.
 This discrepancy is probably due to the
formation of secondary structures in the ssDNA molecule \cite{maier}, 
which can bend back onto itself and form local helices 
where complementary bases A-T and G-C are paired,
gaining an energy of several $kT$ per pair.

The formation of secondary structures is a crucial step in the folding of proteins
and single stranded nucleic acid polymers. Its importance stems from the rather
large values of the binding energy involved in this formation, compared to the
much smaller energy scale of the interaction between secondary structures which
govern the final three-dimensional shape of the molecules (the ternary structure).
As discussed recently \cite{higgs,bunhwa,pagpar}, the formation of
secondary structures in RNA (which is very similar to the one in ssDNA) provides
a wonderful laboratory for detailed studies of some of the basic mechanisms
at work in heteropolymer folding. 

In this paper we modify and extend the previous studies on RNA or ssDNA 
secondary structures 
in order to include one simple aspect of the ternary structure, namely the thermal
fluctuations  of the end-to-end distance, and its dependence on the pulling force. 
Our model, which is solved exactly with generating function techniques,
 provides a detailed description of the elastic 
properties of these polymers. It involves three parameters: the persistence length 
of the molecule, the elastic constant characteristic of bond stretching, and the
pair binding energy.

In the simplest approximation, the backbone of the polymer is 
described by a FJC with $N$  elastic bonds. 
At thermal equilibrium, the probability distribution 
of a bond to be equal to  the vector $\vec r$ is given by 
\be
\mu({\vec r})  = C \exp\((-{( |{\vec r}| -b)^2  \over 2 \ell^2 }\)) \ ,
\label{bond_probability}
\ee
where $b$ is the persistence length and $\ell$ is a length which characterizes
the elasticity of the bond. For RNA or ssDNA, one expects $b$ to be of
the order of a few times the distance between successive bases, and $\ell/b$
to be much smaller than one. 
The spatial conformation  is thus described by the positions 
$vec r_i$, ($i \in\{1,...,N+1\}$) 
of the $N+1$ nodes
which are the articulation points of the chain. 
The attraction between complementary
bases creates an effective  potential $\eps_{ij}(\vec r_i-\vec r_j)$ 
between nodes $i$ and $j$
(arbitrarily far away from each other along the backbone)
which involves a short ranged attraction and a core repulsion. 
We perform a standard virial expansion of the partition function
in terms of the quantities $f_{ij}(\vec r)=\exp(-\eps_{ij}(\vec r)/kT)-1$ which
vanish for $|\vec r|>a$, where $a$ is the range of the interaction. 
The secondary structure is characterized by the set of node pairs $i,j$ such
that $f_{ij} \ne 0$. 

Our main approximation
for describing the secondary structure is the standard one in which one keeps only
 the nested diagrams 
\cite{maccas,zuker,Fontana,dill,higgs,bunhwa,pagpar}, which are defined
as follows:
\begin{itemize}
\item
Each node can be paired to at most one other node.
\item
Two node pairs $i<j$ and $k<l$ (with say $i<k$) 
can coexist only if they are either independent
($i<j<k<l$) or nested ($i<k<l<j$).
\end{itemize}
This condition  neglects the formation
of pseudo-knots, which are known to be rare in RNA folding.  This is thus the
simplest approximation, one in which one adds to the basic elastic model (here for
instance the FJC) the possibility of formation of hairpins, 
consisting of helices, and helices within helices organised in a hierarchical way. 
In the future it should be
possible to generalize our model in order to include some sets of pseudo-knots,
as was done for instance in simulations in \cite{IsaSig}. Including more refined
potentials to get a better description of the secondary structure is also possible
\cite{zuker,Fontana}

\begin{figure}[t!]
\hspace*{-0.5cm}\epsfig{figure=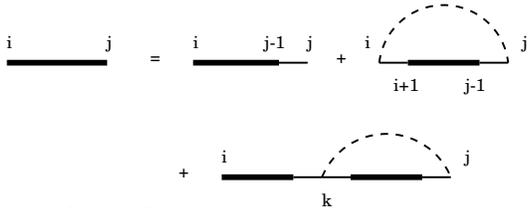,
width=7cm}
\caption{
Recursion relation for the partition function. A thin line denotes one bond,
which in the elastic FJC is a vector chosen with probability
(\ref{bond_probability}). A dashed line between $i$
and $j$ corresponds to an interaction term 
$\exp\((-\eps_{ij}(\vec r)/kT\))-1$. The full line is
the partition function which adds up the effect of all nested or independent
interaction lines.
}
\label{fig_diag}
\end{figure}
The hierarchical structure of the retained diagrams  makes it possible to write
a recursion relation  for the partition function $Z_{j,i}(\vec r)$ 
which describes the
set of nodes $k \in \{ i,i+1,...,j \} $, 
with an end-to-end distance $\vec r_j - \vec r_i = \vec r$.
The recursion is explained in fig. \ref{fig_diag} 
which shows that, when 
$j-i \ge 3$:


\bea
Z_{j,i} (\vec r) &=& \int\! d \vec u\, \mu( \vec u )\, Z_{j-1,i}(\vec r -\vec u )
+f_{ji}(\vec r) \int\!d \vec u_1\,\mu ( \vec u_1)\,
d {\vec u_2}\, \mu( \vec u_2)
Z_{j-1,i+1}(\vec r-\vec u_1-\vec u_2) \\
&+&
\sum_{k=i+1}^{j-2} \int\! \prod_{m=1}^3 d {\vec u_m}\, \mu( \vec u_m)
\, d{\vec v} \,f_{jk}(\vec v) 
Z_{j-1,k+1}(\vec r-\vec u_2-\vec u_3) 
Z_{k-1,i}(\vec r-\vec u_1-\vec v) 
\label{recursion}
\eea


This recursion relation provides the  definition of our model for RNA or 
ssDNA folding. 
It modifies
the  recursions which have been written  previously in the studies of 
RNA secondary structures
\cite{bunhwa} in two  aspects. On the one hand it includes the spatial 
structure, i.e. the positions of the nodes. Secondly it uses the virial expansion
in which the interaction term between $i$ and $j$ is given by $f_{ij}(\vec r)$. 
This is needed
 in order to get back the usual FJC in the limit where the interaction 
potential $\eps$
vanishes. 

As a first step in the study of this model, we investigate in the 
following the case where the interaction energy $\eps_{ij}(\vec r)$ is 
independent of the pair $i,j$. This amounts to using an effective
interaction, averaged over the several bases included within
the persistence length $b$, in which the only effect of the sequence
which is kept is the global concentration in the various base pairs.
This rough approximation  turns out
to be good enough for computing global properties of the polymer like the
force elongation characteristics. The effect of sequence heterogeneities,
which is crucial for dyanmical properties,
is left for future studies.

In the homogeneous case the partition function $Z_{j,i}(\vec r)$ depends only
on $n=j-i$, and is denoted by $Z_n(\vec r)$. 
We introduce the Fourier transform of the  generating function of the $Z_n's$:
\be
\Xi(\zeta,\vec p)=  
\int d \vec r \(( \sum_{n=0}^\infty Z_n (\vec r) \zeta^n\)) 
e^{i \vec p . \vec r} \ ,
\ee
which is expressed in terms of the Fourier transforms:
\bea
\sigma (\vec p) &=& \int d \vec r \  \mu(\vec r) e^{i \vec p . \vec r} \\
f (\vec p) &=& \int d \vec r \ \[[ \exp(-\beta \eps(\vec r))-1\]] 
e^{i \vec p . \vec r} \ .
\eea
Using $Z_0(\vec p) \equiv 1$ and $Z_1(\vec p) \equiv \sigma(\vec p)$
one derives from the recursion relation (\ref{recursion}) the
functional equation:
\be
\Xi(\zeta,\vec p)={1 \over \zeta} {\omega(\zeta,\vec p)  \over 1-
\sigma(\vec p)\omega(\zeta,\vec p)} \ ,
\label{integ_eq}
\ee
where the kernel $\omega$ satisfies the integral equation:
\be
\omega(\zeta,\vec p) = \zeta+\zeta^3 \intq f(\vec p - \vec q)
\sigma(\vec q)^2 \Xi(\zeta,\vec q) \ .
\label{integ_eq_2}
\ee

 The force-elongation characteristics
for a chain with $N$ bonds can be deduced from the partition function
in presence of a force:
\be
Z_N^{\vec F}= \int d \vec r \  Z_N(\vec r) e^{\beta  \vec F . \vec r} \ .
\ee
Its generating function is nothing but $\Xi(\zeta,\vec p_F)$, where
$\vec p_F$ is an imaginary momentum given by $\vec p_F =(0,0,-i \beta F)$
for a force $F$ pulling in the third direction.
For a long chain, $N>>1$, one expects a partition function behaving as 
$Z_N^{\vec F} \sim A \exp(-\beta N \phi(F))/N^\alpha$. The 
free energy per bond $\phi(F)$ determines the radius of convergence of
the series defining the generating function $\Xi(\zeta,\vec p_F)$. 
It is thus equal to 
$ \phi(F) = (1/\beta) \ln (\zeta^*) $,
where $\zeta^*$ is the singularity of $\Xi(\zeta)$
 which is the nearest to the origin.
From the free energy per bond one deduces
 the elongation $L$
along the axis of the force: $L = -N  \partial \phi/ \partial F$,
as well as the average fraction of pairings $n_p$ (defined as the number of
pairings divided by $N$):
$n_p= \partial  \ln (\zeta^*) / \partial \ln (\gamma)$.

The integral equation (\ref{integ_eq}) is easily solved in the case 
where the range of the interaction potential is small compared to $b$
(this approximation is again valid when $b$ is much larger than the 
interbase distance). One can then neglect the momentum dependence of
$f$ and substitute $f(\vec p)$ by the constant
$\gamma b^3$, where $\gamma$ is a dimensionless number 
characteristic of the strength of the pairing and defined by
$\gamma=f(\vec 0)/b^3=
 \int\!d \vec r/b^3 \[[ \exp(-\beta \eps(\vec r))-1\]]$. The kernel
$\omega$ is then momentum independent. 
The relation (\ref{integ_eq_2}) between $\zeta$ and $\omega$ 
can be written as $\omega=\zeta+\zeta^2  A(\omega)$,
where the function $A(\omega)$ is monotonously increasing and such
that $A'(\omega=1)=\infty$. One can then show
 that $\omega(\zeta)$ has a second order branching 
point at  $\zeta_{\rm bp}$ and 
 is analytic for $|\zeta|<\zeta_{\rm bp}$, where $\zeta_{\rm bp}$ is
the maximum of the function $(-1+\sqrt{1+4 \omega A(\omega)})/(2A(\omega))$.

 The singularities of $\Xi$ which control the large $n$ behavior of
$Z_n$ are the branching point of $\omega(\zeta)$ at $\zeta_{\rm bp}$
and the pole at $\zeta_{\rm p}(\vec p)$ determined by the vanishing of the 
denominator of  equation (\ref{integ_eq}), when the momentum is equal to $\vec p_F$:
$
\omega(\zeta_{\rm p}) \sigma( \vec p_F)=1 \ .
$
For purely elastic bonds with $\ell<<b$, one finds 
$
\sigma(\vec p) \simeq \[[ {\sin(pb)/ pb}\]] \exp(-p^2 \ell^2/2) \ , 
$ 
and the pole is located at:
\be
\omega(\zeta_{\rm p})= {\beta F b \over \sinh(\beta F b)}
e^{-\beta^2 F^2 \ell^2/2} \ .
\label{pole_elast}
\ee 

To each of the two singularities is associated one  phase of the model. 
 The ``hairpinned''
phase corresponds to the branching point singularity.
As far as we neglect the momentum dependence of $\omega$ (i.e. for small 
interaction radius) the free energy per bond
$\phi(F) = (1/\beta)\log\zeta_{\rm bp}$ is force independent. The length of the
polymer is thus of order $N^0$ in the long chain ($N\to\infty$) limit.
A fraction $n_{\rm p}$ of  nodes is paired with $n_{\rm p}$ 
independent on the applied force.
The chain is bent in a few, i.e. $O(N^0)$, hairpins, each one involving $O(N)$
bonds.
The ``elongated'' phase corresponds to the pole singularity. 
The free energy   $\phi(F) = (1/\beta)\log\zeta_{\rm p}(\vec p_F)$ is force 
dependent and the elongation is extensive (proportional to $N$). 
This can be written as: 
$L(F) = n_{\rm free}(F) L_{FJC}(F)$ where $L_{FJC}(F)$ is the elongation without 
interaction (i.e. in the case $\gamma=0$) 
and $n_{\rm free}(F)$ the fraction of nodes which do not belong to
any hairpin. The fraction of pairings rapidly decreases with the applied 
force. The number of hairpins is  $O(N)$.

In our model there exists a second order phase transition
between the hairpinned phase
at low force  and the elongated phase at  high 
force. This phase transition is a robust feature of the model 
which does not depend on the details of the interaction potential and of the
bond stretching potential: the branched point singularity, 
associated with the hairpinned phase, is present 
as soon as $\sigma(\vec p) \sim 1- \kappa \vec p ^2$ for
small $|\vec p|$,
which is the generic situation. The pole singularity, associated with the
elongated phase, is always present.
 The boundary between the two phases  occurs at a critical 
force $F_c(\gamma)$ which 
increases monotonically with $\gamma$. 
Slightly above the threshold the
elongation grows linearly with the force: $L(F)\propto F-F_c(\gamma)$.
The asymptotic behaviors of the  dimensionless critical force
are:
\bea
\beta b F_c(\gamma) & \sim & \frac{1}{4}\log ( \gamma)  
\quad\quad\mbox{for}\quad
 1\ll \gamma \ll e^{b^2/\ell^2}\\
\beta b F_c(\gamma) & \sim & \frac{\gamma}{8\pi\kappa^2} \quad\mbox{for}\quad
\gamma\ll 1
\eea
Notice that the 
linear dependence of $F_c$ at small $\gamma$
a prediction which is independent of the detailed form of the bond probability 
distribution (\ref{bond_probability}). 

Equations (\ref{integ_eq_2}) and (\ref{pole_elast}) 
can be easily solved numerically. We compared our theoretical predictions
with the experimental data presented in Ref. \cite{maier}
on the 
$\ln$ force versus elongation characteristic for a
charomid-ssDNA at room temperature under different salinity conditions.
Using the elastic model for bond stretching (\ref{bond_probability}),
our three fitting parameters are:
the persistence length $b$, the elasticity $\ell$, the interaction parameter 
$\gamma$. 
\begin{figure}[t!]
\hspace*{-0.5cm}\epsfig{figure=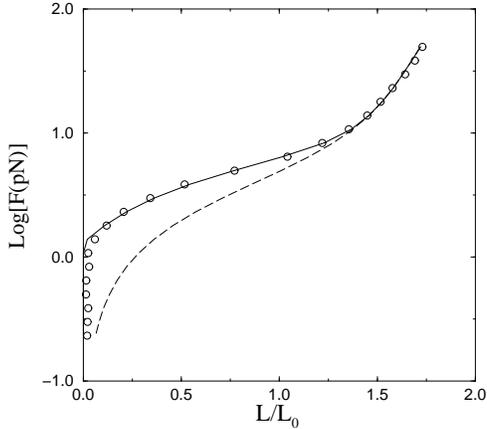,
 angle=-90,width = 7cm}
\caption{ Fit of the force-elongation characteristics of charomid ssDNA.
The circles are the experimetal data of Ref [12]. 
The continuous line is the best fitting curve obtained with our model model. 
The dashed curve is the FJC characteristics obtained by switching off the 
interaction.
The difference between the two is due to the formation of hairpins.}
\label{fit}
\end{figure}
As shown in fig.\ref{fit}, we obtain a good agreement with the experimental curve
at the highest salt concentration
($10$ mM PB, $5$ mM Mg).
The small elongation region ($L/L_0<.1$) of the experiment was not considered
since the interactions of the molecule with the glass plate cannot be ignored 
(this forbids a  study of the critical force region
with the present data).  The fitting parameters are:
the persistence length $b$, the elasticity $\ell$, the interaction parameter 
$\gamma$. The number of bonds was fixed as
in \cite{maier} such that $N b = 1.6875\, L_0$,
where $L_0$ is the crystallographic length of the double stranded DNA.
A least square fit,
 yields the following results:
$b = 19.2\,{\rm \AA}$,   
$\gamma = 1.89$, $\ell = 1.01\,{\rm \AA}$.
The orders of magnitudes of the various parameters are correct.
The persistence length is of the order of three times the interbase distance
$b_0$ (our approximation of a large value of $b>b_0$ 
 is marginally self-consistent and should be improved upon
in the future). The value of $\ell$,
when expressed in terms of the
enthalpic elasticity $S$ as in \cite{maier},
 corresponds to 
$S=b/(\beta \ell^2)\simeq 1000 \, pN$, typical of the values
measured at higher forces\cite{smith2,heslot,gaub}; the 
 approximation $\ell <<b$ is valid.
The value of $\gamma$ is characteristic of the strength of the interaction.
For a potential well of width $a$ and depth $\eps_0$, one has
$\gamma \sim (a/b)^3 \exp(\beta \eps_0)$, which is compatible
with some typical values such as $a \sim 4\, { \rm \AA}$,
$\eps_0 \sim 2.9\, kT$.
\begin{figure}[t!]
\hspace*{-0.5cm}\epsfig{figure=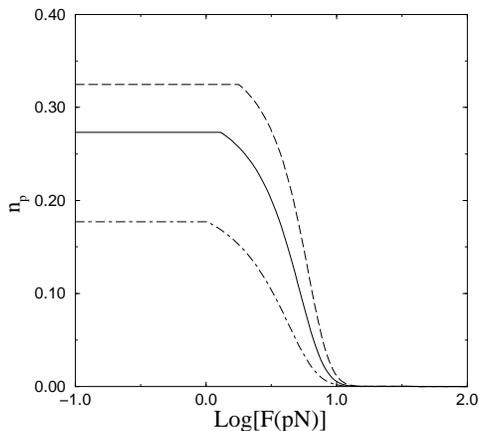,angle=-90,width=7cm}
\caption{The fraction of paired nodes in the secondary structure
 as a function of the external pulling force. The three curves refer to three 
different values of the interaction parameter $\gamma$. From top to bottom
$\gamma = 3.9,\, 1.9, \, 0.59$. The other parameters of the model
correspond to the experimental situation: they are 
 fixed as in the fit of Fig. \ref{fit}.}
\label{pairs}
\end{figure}
From our computation one can deduce the pairing
fraction $n_p$ in the conditions of the experiment. This is plotted
in Fig. \ref{pairs}. 
It is clear that in the region of forces above $10\, pN$ there is no pairing.
This is consistent with the measurements of ref. \cite{maier} 
which showed that the charcteristics of two different 
ssDNA's with different G-C concentration merge in that region.
One should keep in mind that  the  two fitting parameters
$b$ and $\ell$ are basically fixed by this high force region
where there is no pairing.
The low force part with pairing is the one 
which fixes the binding parameter $\gamma$.

When the salinity is lowered some new physical effects become relevant.
The electrostatic interactions between the bases are less effectively screened,
and probably  the FJC is not  a good model. Our model 
should still give a reasonable account of the effect of secondary structures 
on the elongation. One possibility to test it is to use, instead of the elastic FJC,
the experimental force elongation characteristics measured on a molecule 
exposed first to  a chemical treatment (for instance glyoxal) which destroys
the ability of the bases to pair. Our model should then allow to deduce from
two experimental curves (one with glyoxal, the other without) the effect of the
secondary structures \cite{all}. 

In this paper we have introduced a solvable model of the structure of ssDNA
or RNA
molecules which includes, together with the secondary structure,
 one important element of its ternary structure. 
The model gives a general framework for including the effect of 
hairpin formation in the elongation
properties of the molecules.
When used with a simple FJC model for the polymer 
without hairpins, it is in good agreement with the experimental data 
at high ionic concentration. Several extensions of this study are
natural. The  description of data obtained at smaller
ionic concentration requires to go beyond the FJC approximation. Another 
natural extension of our study, also possible within this model,
 is to study the effects of the disorder in the sequence of bases.

Acknowledgements: It is a pleasure to thank D. Bensimon and
V. Croquette for many useful discussions and for providing us their data.


\end{document}